\documentclass[aps,prl,reprint,a4paper,superscriptaddress,showkeys,floatfix]{revtex4-1}
\usepackage[utf8]{inputenc}              %  Determina a codificação utilizada
\usepackage[english]{babel}              %  Língua para formatação do texto.
                                         
\usepackage[T1]{fontenc}                 %  Seleção de códigos de fonte.

\usepackage{lmodern}                     %  Usa a fonte Latin Modern

\usepackage{microtype}                   %  Melhora a justificação - deve ser
\usepackage{xcolor}                      %  Para cores

\usepackage{graphicx}                    %  Para inclusão de figuras (png, jpg, gif, bmp, pdf)

\usepackage[bf]{subfigure}

\usepackage{amssymb,amsmath,amsfonts}      %  Símbolos matemáticos da AMS

\usepackage{mathtools}

\usepackage{braket}                        %  Notação braket

\usepackage{tensor}                        %  Representação de tensores

\usepackage[smaller]{cancel}        %  Desenha setas em modo matemático para substituição de valores.

\usepackage{upgreek}     % Para ter mais variações de letras gregas
\usepackage{bm}          % bold symbols
\usepackage[colorlinks,citecolor=blue,filecolor=black,linkcolor=red,urlcolor=blue]{hyperref}
\newcommand{\Pb}[2]{\left\{#1,#2\right\}_{P}}    %  Parênteses de Poisson.
\newcommand{\Co}[2]{\left[#1,#2\right]_{-}}      %  Comutador.
\newcommand{\bs}[1]{\boldsymbol{#1}}
\begin{document}

\title{On a proper-time approach to localization}

\date{\today}

\author{E. R. F. Taillebois}
\email{emile.taillebois@ifgoiano.edu.br}
%\affiliation{Instituto de Física, Universidade Federal de Goiás, 74.690-900, Goiânia, Goiás, Brazil}
\affiliation{Instituto Federal Goiano - Campus Avançado Ipameri, 75.780-000, Ipameri, Goiás, Brazil}

\author{A. T. Avelar}
\affiliation{Instituto de Física, Universidade Federal de Goiás, 74.690-900, Goiânia,
Goiás, Brazil}

\begin{abstract}
The causality issues concerning Hegerfeldt's paradox and the localization of relativistic quantum systems are addressed through a proper-time formalism of single-particle operators. The proposed description does not depend on classical parameters connected to an external observer and results in a single-particle formalism in which localization is described by explicitly covariant four-vector operators associated with POVM measurements parametrized by the system's proper-time. As a consequence, it is shown that physically acceptable states are necessarily associated with the existence of a temporal uncertainty and their proper-time evolution is not subject to the causality violation predicted by Hegerfeldt.
\end{abstract}



\maketitle

\emph{Introduction -} The introduction of relativistic effects in quantum information theory, an ongoing program that has its origins in the founding works of Czachor \cite{Czachor1997} and Peres et al. \cite{Peres2002}, was responsible for attracting new attention to problems such as the proper definition of the concept of relativistic spin \cite{Czachor2003, Peres2004, Saldanha2012, Taillebois2013, Palmer2013, Giacomini2019, Taillebois2020} and the causality issue concerning localization in relativistic quantum systems \cite{Terno2014,Celeri2016,Silva2019}. Both these problems are intimately connected, as it has long been known that different notions of localization lead to distinct concepts of relativistic spin \cite{Pryce1948, Fleming1965a}. In addition, the localization problem may also affects certain predictions related to quantum information theory, since data-processing is typically performed in limited space-time intervals \cite{Palmer2012,Caban2014}.

The first evidence concerning the incompatibility of localization and causality has its origins in Hegerfeldt's well known theorem \cite{Hegerfeldt1974, Hegerfeldt1980}, which established the causality violation for initially strictly localized states with a well defined energy sign. Since this directly affects any attempt to describe a position detection in terms of self-adjoint (s.a.) operators in the context of relativistic quantum mechanics (RQM), the remaining alternative was the use of POVMs approaches to localization \cite{Terno2014,Celeri2016}. However, even those may be subject to causality issues, as Hegerfeldt's theorem was later extended to states with an exponentially bounded decay \cite{Hegerfeldt1985}.

In this letter it is shown that, when a proper-time approach is applied to the description of a single-particle system in RQM, any physical acceptable state will present an inherent temporal uncertainty. Together with the requirement of covariance, this leads to a natural POVM description of localization with a proper-time evolution for which Hegerfeldt's results do not apply. The investigated model is that of a free spinless particle of fixed mass $m$, a choice justified by the need to adequately underpin the resulting notion of localization before considering any potential contribution from the introduction of internal degrees of freedom. Complete derivations and complementary discussions are presented in the companion paper \cite{Taillebois2020pra}.

Starting from a classical description, the particle's four-momentum $\Pi^{\mu}$ and angular momentum tensor $J^{\mu\nu}$ can be used to describe the system's world line by means of the four coordinates
\begin{equation}
Q^{\mu}(\tau) = -\frac{J^{\mu\lambda}\Pi_{\lambda}}{m^{2}} + \frac{\Pi^{\mu}}{m}\tau, \label{eq:NewQt}
\end{equation}
with the particle's proper-time $\tau$ acting as parameter. On the constant mass surface $\Pi^{\mu}\Pi_{\mu} + m^{2} = 0$, those quantities satisfy the relationship $Q^{\mu}(\tau)\Pi_{\mu} = -m\tau$ and the Poisson brackets
\begin{equation}
\Pb{J^{\mu\nu}}{Q^{\sigma}(\tau)} = 2\eta^{\sigma[\mu}Q^{\nu]}(\tau), \label{eq:PosPar}
\end{equation}
which ensure that the coordinates in \eqref{eq:NewQt} form a four-vector quantity interpreted as the system's four-position with explicit proper-time parameterization.

The direct quantization of the quantities $Q^{\mu}(\tau)$ over a physical Hilbert space $\mathcal{H}_{phys}$ using only the correspondence principle as a guiding resource is not a simple task due to the Poisson brackets
\begin{subequations}
\begin{eqnarray}
\Pb{Q^{\mu}(\tau)}{Q^{\nu}(\tau)} = \frac{J^{\mu\nu}}{m^{2}}, \nonumber \\
\Pb{Q^{\mu}(\tau)}{\Pi_{\nu}} = \tensor[]{\eta}{^{\mu}_{\nu}} + \frac{\Pi^{\mu}\Pi_{\nu}}{m^{2}}, \nonumber
\end{eqnarray}
\end{subequations}
valid over the constant mass surface. However, using Dirac's quantization approach \cite{Gitman1990} and employing the so-called group averaging technique \cite{Marolf1995(1),Marolf1995(2),Ashtekar1995,Louko2006}, the quantization of those quantities can be achieved in an unambiguous way, resulting in the physical Hilbert space $\mathcal{H}_{phys} = \mathcal{H}_{phys}^{+} \oplus \mathcal{H}_{phys}^{-} = L^{2}(\mathbb{R}^{3},d\mu(\bs{\pi}))\oplus L^{2}(\mathbb{R}^{3},d\mu(\bs{\pi}))$, with $d\mu(\bs{\pi}) = md^{3}\pi/E_{\bs{\pi}}$ and $E_{\bs{\pi}} = \sqrt{\|\bs{\pi}\|^2 + m^{2}}$. Over this space, the symmetrized form of the quantities $Q^{\mu}(\tau)$ give rise to the acting rules
\begin{eqnarray*}
\check{Q}_{phys}^{0}(\tau) & = & \sigma^{3}\frac{E_{\bs{\pi}}}{m}\left[\frac{i}{m}\left(\bs{\pi}\cdot\nabla_{\bs{\pi}} + \frac{3}{2}\right) + \tau\right], \\
\check{Q}_{phys}^{j}(\tau) & = & \sigma^{3}\left[i\left(\frac{\partial}{\partial \pi_{j}} + \frac{\pi^{j}}{m^2}\bs{\pi}\cdot\nabla_{\bs{\pi}} + \frac{3}{2}\frac{\pi^{j}}{m^2}\right) +\frac{\pi^{j}}{m}\tau\right],
\end{eqnarray*}
while the momentum variables $\Pi^{\mu}$ and the angular momentum tensor $J^{\mu\nu}$ lead to the following operations:
\begin{eqnarray*}
\check{\Pi}_{phys}^{\mu} & = & \sigma^{3}\left(E_{\bs{\pi}} \tensor{\eta}{^{\mu}_{0}} + \pi^{j}\tensor{\eta}{^{\mu}_{j}}\right), \\
\check{J}_{phys}^{\mu k} & = & -\tensor{\eta}{^{\mu}_{0}}iE_{\bs{\pi}}\frac{\partial}{\partial \pi_{k}} + \tensor{\eta}{^{\mu}_{j}}i\left(\pi^{k}\frac{\partial}{\partial \pi_{j}} -\pi^{j}\frac{\partial}{\partial \pi_{k}}\right).
\end{eqnarray*}

The operators $\hat{Q}^{\mu}_{phys}(\tau)$ are interpreted as the proper-time parameterized coordinates of the system's four-position, the commutation relations
\begin{equation}
\Co{\hat{J}^{\mu\nu}_{phys}}{\hat{Q}^{\sigma}_{phys}(\tau)} = 2i\eta^{\sigma[\mu}\hat {Q}^{\nu]}_{phys}(\tau) \label{eq:ASD}
\end{equation}
ensuring the Lorentz four-vector character and the covariant aspect of the related localization concept. Besides \eqref{eq:ASD}, the additional commutation relations
\begin{eqnarray*}
\Co{\hat{Q}^{\mu}_{phys}(\tau)}{\hat{Q}^{\nu}_{phys}(\tau)} & = & i\frac{\hat{J}^{\mu\nu}_{phys}}{m^2}, \\
\Co{\hat{Q}^{\mu}_{phys}(\tau)}{\hat{\Pi}^{\nu}_{phys}} & = & i\left(\tensor[]{\eta}{^{\mu\nu}} + \frac{\hat{\Pi}^{\mu}_{phys}\hat{\Pi}^{\nu}_{phys}}{m^2}\right)
\end{eqnarray*}
imply that the four-position components are non-commutative and, together with $\hat{J}^{\mu\nu}_{phys}$, satisfy a deSitter-like algebra with fundamental length $1/m$ \cite{Snyder1947,Aldrovandi2016}, the usual canonical behavior being recovered in the non-relativistic limit.

Since $\check{\Pi}^{\mu}_{phys}$ corresponds to the direct sum of multiplicative operations, the s.a. character of the operator $\hat{\Pi}^{\mu}_{phys}$ can be immediately stated by assuming the natural domains $D_{\Pi_{\mu}} = \left\{\phi|\phi(\bs{\pi}), \sigma^{3}\left(E_{\bs{\pi}} \tensor{\eta}{^{\mu}_{0}} + \pi^{j}\tensor{\eta}{^{\mu}_{j}}\right)\phi(\bs{\pi}) \in \mathcal{H}_{phys}\right\}$. On the other hand, in the case of operators $\hat{Q}^{\mu}_{phys}(\tau)$, it is necessary to calculate their so-called deficiency indices \cite{Gitman2012}, since these will generally be unbounded operators.

\emph{Self-adjoint extensions of $\hat{Q}^{0}_{phys}(\tau)$ -} The operators $\hat{Q}^{0}_{phys}(\tau)$, $\hat{J}^{12}_{phys}$ and $\|\hat{\mathbf{J}}_{phys}\|^{2}$ form a complete set of commuting observables (CSCO), the s.a. character of $\hat{J}^{12}_{phys}$ and $\|\hat{J}_{phys}\|^{2}$ being verifiable as in the usual non-relativistic scenario and leading to the same spectrum and corresponding eigenstates. On the other hand, for $\hat{Q}^{0}_{phys}(\tau)$, the complete definition of the operator requires a detailed analysis of its deficiency indices and domain.

Using spherical coordinates and adopting $C^{\infty}_{0}(\mathbb{R}^3)\oplus C^{\infty}_{0}(\mathbb{R}^3)$ as initial domain for the operator $\hat{Q}^{0}_{phys}(\tau)$, the domain of the corresponding adjoint $\hat{Q}^{0*}_{phys}(\tau)$ is given by
\begin{widetext}
\begin{equation}
D_{\hat{Q}^{0}_{phys}(\tau)}^{*} = \left\{\phi_{*}(r_{\pi})\,|\, r_{\pi}E_{r_{\pi}}\phi_{*}(r_{\pi}) \text{ is a.c. in }\mathbb{R}_{\geq 0}; \; \phi_{*}(r_{\pi}),\check{Q}^{0}_{phys}(\tau)\phi_{*}(r_{\pi}) \in L^{2}(\mathbb{R}_{\geq 0}, d\mu(r_{\pi}))\oplus L^{2}(\mathbb{R}_{\geq 0}, d\mu(r_{\pi}))\right\}, \nonumber
\end{equation} 
\end{widetext}
where  $d\mu(r_{\pi}) = \frac{mr_{\pi}^{2}}{E_{r_{\pi}}}dr_{\pi}$ and a.c. stands for "absolute continuous". Thus, solving $\check{Q}^{0}_{phys}(\tau)R^{t}_{\tau}(r_{\pi}) = tR^{t}_{\tau}(r_{\pi})$ for $t = \pm i/m$, it results that the operator $\hat{Q}^{0}_{phys}(\tau)$ has deficiency indices $\eta = (1,1)$ and an infinite number of s.a. extensions with a single parameter, since both solutions
\begin{equation}
R^{\xi i/m}_{\tau}(r_{\pi}) =
\frac{\sqrt{2}e^{im\tau \ln (\frac{r_{\pi}}{m})}}{r_{\pi}^{1/2}(E_{r_{\pi}} + m)}
\begin{pmatrix}
\delta_{\xi +} \\
\delta_{\xi -}
\end{pmatrix}, \nonumber
\end{equation}
with $\xi = \pm$, belong to $D_{\hat{Q}^{0}_{phys}(\tau)}^{*}$. It is important to note that the projections of $\hat{Q}^{0}_{phys}(\tau)$ over the subspaces $\mathcal{H}_{phys}^{\pm}$ do not have s.a. extensions, since they have deficiency indexes given by $\eta = (1,0)$ (positive energy projection) and $\eta = (0,1)$ (negative energy projection). This implies that an eigenfunction of a s.a. extension of $\hat{Q}^{0}_{phys}(\tau)$ cannot have a well defined energy sign and, even more importantly, the nonexistence of single-particle s.a. extension for $\hat{Q}^{0}_{phys}(\tau)$. 

To define the s.a. extensions of the symmetric operator $\hat{Q}^{0}_{phys}(\tau)$ it is necessary first to write its closure $\underline{\hat{Q}^{0}_{phys}(\tau)}$. Using the sesquilinear form
\begin{equation}
w_{*}(\phi_{*},\psi_{*}) = (\phi_{*}, \hat{Q}^{0*}_{phys}\psi_{*})_{phys} - (\hat{Q}^{0*}_{phys}\phi_{*}, \psi_{*})_{phys}, \nonumber
\end{equation}
the closed operator $\underline{\hat{Q}^{0}_{phys}(\tau)}$ is given by the acting rule $\underline{\hat{Q}^{0}_{phys}(\tau)}\underline{\phi} = \hat{Q}^{0*}_{phys}(\tau) \underline{\phi}$ defined over the domain
\begin{equation}
D_{\underline{\hat{Q}^{0}_{phys}(\tau)}} = \left\{ \underline{\phi} \Big|\, \underline{\phi} \in D_{\hat{Q}^{0}_{phys}(\tau)}^{*};\; w_{*}(\underline{\phi},R^{\pm i/m}_{\tau}(r_{\pi})) = 0\right\}, \nonumber
\end{equation}
which implies the boundary condition $\lim_{r_{\pi}\rightarrow\infty}|\underline{\phi_{\xi}(r_{\pi})}| < \lim_{r_{\pi}\rightarrow\infty}\mathcal{O}(r_{\pi}^{-3/2}) = 0$. Then, from the Main Theorem of s.a. extensions \cite{Gitman2012}, the one-parameter family of s.a. extensions $\tensor[_{\varphi}]{\hat{Q}}{^{0}_{\hspace{-0.1cm}phys}}(\tau)$, with parameter $\varphi \in (-\uppi,\uppi]$, can be defined as
\begin{widetext}
\begin{equation}
\tensor[_{\varphi}]{\hat{Q}}{^{3}_{\hspace{-0.1cm}phys}}(\tau)\,:\, \left\{
\begin{aligned}
& D_{\hspace{-0.15cm}\tensor[_{\varphi}]{\hat{Q}}{^{0}_{\hspace{-0.1cm}phys}}(\tau)} = \left\{ \phi_{\varphi} \in D_{\hat{Q}^{0}_{phys}(\tau)}^{*} \,\Big|\, \right. w_{*}(R^{+i/m}_{\tau} + e^{i\varphi}R^{-i/m}_{\tau},\phi_{\varphi}) = 0\bigg\} \\
& \tensor[_{\varphi}]{\hat{Q}}{^{0}_{\hspace{-0.1cm}phys}}(\tau)\phi_{\varphi} = \hat{Q}^{0*}_{phys}(\tau) \phi_{\varphi}
\end{aligned}\right., \nonumber
\end{equation}
\end{widetext}
which implies the boundary condition $\lim_{r_{\pi} \rightarrow \infty} \left[\phi_{\varphi;+}(r_{\pi}) - e^{-i\varphi}\phi_{\varphi;-}(r_{\pi})\right] = 0$ with a decay at infinity that must be bounded according to $|\phi_{\varphi;+}(r_{\pi}) - e^{-i\varphi}\phi_{\varphi;-}(r_{\pi})| < \mathcal{O}(r_{\pi}^{-3/2})$.

The spectrum of the s.a. extension $\tensor[_{\varphi}]{\hat{Q}}{^{0}_{\hspace{-0.1cm}phys}}(\tau)$ does not have a discrete component since there exist no $R^{t}(r_{\pi};\tau) \in L^{2}(\mathbb{R}_{\geq 0},d\mu(r_{\pi}))\oplus L^{2}(\mathbb{R}_{\geq 0},d\mu(r_{\pi}))$ such that $\check{Q}^{0}_{phys}(\tau)R^{t}(r_{\pi};\tau) = tR^{t}(r_{\pi};\tau)$ for $t\in \mathbb{R}$. However, since $(\tensor[_{\varphi}]{\hat{Q}}{^{0}_{\hspace{-0.1cm}phys}}(\tau) -t\hat{I})^{-1}$ exists and is unbounded for all $t \in \mathbb{R}$, this s.a. extension has a continuous spectrum $t \in \mathbb{R}$ with a complete set of orthogonal eigenfunctions given by
\begin{equation}
R^{t}_{\varphi}(r_{\pi},\tau) = \sqrt{\frac{m}{2\uppi}}r_{\pi}^{-3/2}\left(\frac{r_{\pi}}{m}\right)^{im\tau}
\begin{pmatrix}
\left(\frac{r_{\pi}}{E_{r_{\pi}} + m}\right)^{-imt} \\
e^{i\varphi}\left(\frac{r_{\pi}}{E_{r_{\pi}} + m}\right)^{imt}
\end{pmatrix}. \nonumber
\end{equation}
Thus, the complete set of generalized orthogonal eigenfunctions associated to the present CSCO is given by $\psi^{t,l,m_{z}}_{\varphi}(\bs{\pi};\tau) = Y^{l,m_{z}}(\Omega_{\pi})R^{t}_{\varphi}(r;\tau)$, where $Y^{l,m_{z}}(\Omega_{\pi})$ are the spherical harmonics with $l\in\mathbb{N}_{0}$ and $-l \leq m_{z} \leq l$.

\emph{Self-adjoint extensions of $\hat{Q}^{3}_{phys}(\tau)$ -} A CSCO for $\hat{Q}^{3}_{phys}(\tau)$ can be obtained using the operators $\hat{J}^{12}_{phys}$ and $\hat{O}_{phys} \equiv (\hat{J}_{phys}^{12})^{2} - (\hat{J}_{phys}^{01})^2 - (\hat{J}_{phys}^{02})^2$.

Adopting the hyperbolic coordinates $\omega_{\pi} \in [0,\infty)$, $\nu_{\pi} \in (-\uppi/2,\uppi/2)$ and $\varphi_{\uppi} \in [0,2\uppi)$, related to the original Cartesian coordinates through $\pi^{1} = m\sinh(\omega_{\pi})\sec(\nu_{\pi})\cos(\varphi_{\pi})$, $\pi^{2} = m\sinh(\omega_{\pi})\sec(\nu_{\pi})\sin(\varphi_{\pi})$ and $\pi^{3} = m\tan(\nu_{\pi})$, the acting rule of $\hat{Q}^{3}_{phys}(\tau)$ may be written as
\begin{equation}
\check{Q}^{3}_{phys}(\tau) = \left[\frac{i}{m}\left(\frac{\partial}{\partial \nu_{\pi}} + \frac{3}{2}\tan(\nu_{\pi}) \right) + \tan(\nu_{\pi})\tau\right]\sigma^{3}. \nonumber
\end{equation}
Thus, starting with $C_{0}^{\infty}(-\uppi/2,\uppi/2)\oplus C_{0}^{\infty}(-\uppi/2,\uppi/2)$ as an initial domain for the definition of $\hat{Q}^{3}_{phys}(\tau)$, the corresponding adjoint $\hat{Q}^{3*}_{phys}(\tau)$ is completely defined by its domain
\begin{widetext}
\begin{eqnarray*}
D^{*}_{\hat{Q}^{3}_{phys}(\tau)} =  \Big\{\phi_{*}(\nu_{\pi}) & \,|\, & \phi_{*}(\nu_{\pi}) \text{ is a.c. in } [-\pi/2,\pi/2]; \\ & & \phi_{*}(\nu_{\pi}),\check{Q}^{3}_{phys}(\tau)\phi_{*}(\nu_{\pi}) \in L^{2}((-\pi/2,\pi/2), d\mu(\nu_{\pi}))\oplus L^{2}((-\pi/2,\pi/2), d\mu(\nu_{\pi})) \Big\},
\end{eqnarray*}
\end{widetext}
where $d\mu(\nu_{\pi}) = \sec^{3}(\nu_{\pi})d\nu_{\pi}$. Solving $\check{Q}^{3}_{phys}(\tau)\mathcal{V}^{z}_{\tau}(\nu_{\pi}) = z\mathcal{V}^{z}_{\tau}(\nu_{\pi})$ for $z = \pm i/m$, it results that all the solutions 
\begin{equation}
\mathcal{V}^{\pm i/m}_{(\xi);\tau}(\nu_{\pi}) = \frac{1}{\sqrt{\sinh(\uppi)}}\frac{(\sec\nu_{\pi})^{im\tau}}{(\sec\nu_{\pi})^{3/2}}e^{\pm \xi\nu_{\pi}}\begin{pmatrix} \delta_{\xi +} \\ \delta_{\xi -} \end{pmatrix},
\end{equation}
with $\xi = \pm$, belong to $D^{*}_{\hat{Q}^{3}_{phys}(\tau)}$. Thus, the operator $\hat{Q}^{3}_{phys}(\tau)$ has deficiency indices $\eta = (2,2)$ and an infinite number of s.a. extensions parameterized by four parameters $\varphi = \{\varphi_{n};\; n = 1,2,3,4;\; \varphi_{n} \in (-\uppi,\uppi]\}$. Unlike what was found for $\hat{Q}^{0}_{phys}(\tau)$, the projections of $\hat{Q}^{3}_{phys}(\tau)$ over the subspaces $\mathcal{H}_{phys}^{\pm}$ also have s.a. extensions, since their deficiency indices are given by $\eta_{\pm} = (1,1)$.

The closure $\underline{\hat{Q}^{3}_{phys}(\tau)}$ is given by the acting rule $\underline{\hat{Q}^{3}_{phys}(\tau)}\underline{\phi} = \hat{Q}^{3*}_{phys}(\tau)\underline{\phi}$ defined over the domain
\begin{equation}
D_{\underline{\hat{Q}^{3}_{phys}(\tau)}} = \left\{\underline{\phi} \,|\, \underline{\phi} \in D^{*}_{\hat{Q}^{3}_{phys}(\tau)}; \, w_{*}\left(\underline{\phi},\mathcal{V}^{\pm i/m}_{(\xi);\tau}\right) = 0 \right\}, \nonumber
\end{equation}
the requirement $w_{*}\left(\underline{\phi},\mathcal{V}^{\pm i/m}_{(\xi);\tau}\right) = 0$ being equivalent to the boundary condition
\begin{equation}
\lim_{\nu_{\pi}\rightarrow \pm \frac{\pi}{2}}\underline{\phi_{\xi}(\nu_{\pi})} = 0 \label{eq:eqh}
\end{equation}
with a decay bounded by $\mathcal{O}((\sec\nu_{\pi})^{-3/2})$ for $\nu_{\pi}\rightarrow\pm\uppi/2$. Then, denoting by $\mathfrak{N}_{\mp i/m}$ the space spanned by $\mathcal{V}^{\pm i/m}_{(\xi);\tau}(\nu_{\pi})$, the parameterized s.a. extensions of $\hat{Q}^{3}_{phys}(\tau)$ can be defined as
\begin{widetext}
\begin{equation}
\tensor[_{\varphi}]{\hat{Q}}{^{3}_{\hspace{-0.1cm}phys}}(\tau)\,:\, \left\{
\begin{aligned}
D_{\tensor[_{\varphi}]{\hat{Q}}{^{3}_{\hspace{-0.1cm}phys}}(\tau)} & = \Big\{\phi_{\varphi} \in D^{*}_{\hat{Q}^{3}_{phys}(\tau)} \,\Big|\, w_{*}\left(\phi_{i/m} + \hat{U}(\varphi)\phi_{i/m}, \phi_{\varphi}\right) = 0, \;\forall \phi_{i/m} \in \mathfrak{N}_{-i/m}\Big\} \\
\tensor[_{\varphi}]{\hat{Q}}{^{3}_{\hspace{-0.1cm}phys}}(\tau)&\phi_{\varphi} = \underline{\hat{Q}^{3}_{phys}(\tau)}\underline{\phi} + \frac{i}{m}\phi_{i/m} - \frac{i}{m}\hat{U}(\varphi)\phi_{i/m} = \hat{Q}^{3*}_{phys}(\tau)\phi_{\varphi}
\end{aligned}\right., \nonumber
\end{equation}
\end{widetext}
where the isometric map $\hat{U}(\varphi)$ is given by
\begin{equation}
\begin{aligned}
\hat{U}(\varphi) : \mathfrak{N}_{-i/m} & \rightarrow \mathfrak{N}_{+i/m} \\
\mathcal{V}^{i/m}_{(\xi);\tau}(\nu_{\pi}) & \mapsto \hat{U}(\varphi)\mathcal{V}^{i/m}_{(\xi);\tau}(\nu_{\pi}) = \hspace{-1pt}\sum_{\xi^{\prime}}U_{\xi^{\prime}\xi}(\varphi)\mathcal{V}^{-i/m}_{(\xi^{\prime});\tau}(\nu_{\pi})
\end{aligned} \nonumber
\end{equation}
with factors $U_{\xi^{\prime}\xi}(\varphi)$ forming an arbitrary $U(2)$ matrix:
\begin{equation}
U(\varphi) \;\dot{=}\; e^{i\varphi_{1}}
\begin{pmatrix}
e^{i\varphi_{2}}\cos(\varphi_{4})   & e^{i\varphi_{3}}\sin(\varphi_{4})  \\
-e^{-i\varphi_{3}}\sin(\varphi_{4}) & e^{-i\varphi_{2}}\cos(\varphi_{4})
\end{pmatrix}.\nonumber
\end{equation}

The general form of the $U(2)$ matrices allows s.a. extensions $\tensor[_{\varphi}]{\hat{Q}}{^{3}_{\hspace{-0.1cm}phys}}(\tau)$ that are not necessarily single-particle observables. However, single-particle s.a. extensions can be constructed using the direct sum of the s.a. extensions associated with the projections $\hat{Q}^{3,\pm}_{phys}(\tau)$ over the subspaces $\mathcal{H}_{phys}^{\pm}$. These extensions correspond to the operators $\tensor[_{\varphi}]{\hat{Q}}{^{3}_{\hspace{-0.1cm}phys}}(\tau)$ with $\varphi_{2}=\varphi_{4}=0$ and, from now on, will be the extensions of interest. 

The well-defined energy sign s.a. extensions $\tensor[_{\varphi}]{\hat{Q}}{^{3,\xi}_{\hspace{-0.4cm}phys}}(\tau)$, with $\xi = \pm$, are given by the acting rule $\tensor[_{\varphi}]{\hat{Q}}{^{3,\xi}_{\hspace{-0.4cm}phys}}(\tau)\phi_{\varphi} = \hat{Q}^{3*,\xi}_{phys}(\tau)\phi_{\varphi}$ with domain $D_{\tensor[_{\varphi}]{\hat{Q}}{^{3,\xi}_{\hspace{-0.4cm}phys}}(\tau)}$ given by
\begin{equation}
 \left\{\phi_{\varphi} \in D^{*}_{\hat{Q}^{3,\xi}_{phys}(\tau)} \,|\, w_{*}\left(\mathcal{V}^{i/m}_{(\xi);\tau} + e^{i\varphi}\mathcal{V}^{-i/m}_{(\xi);\tau}, \phi_{\varphi}\right) = 0\right\}, \nonumber
\end{equation}
the imposed restriction leading to the boundary condition
\begin{widetext}
\begin{equation}
 \lim_{\nu_{\pi}\rightarrow\frac{\pi}{2}}\left[\left(e^{\xi\uppi/2} + e^{-i\varphi}e^{-\xi\uppi/2}\right)\phi_{\varphi}(\nu_{\pi})
 -  \left(e^{-\xi\uppi/2} + e^{-i\varphi}e^{\xi\uppi/2}\right)\phi_{\varphi}(-\nu_{\pi})\right] < \lim_{\nu_{\pi}\rightarrow\frac{\pi}{2}}\mathcal{O}\left((\sec\nu_{\pi})^{-3/2}\right) = 0. \label{eq:Bound1}
\end{equation}
\end{widetext}
Thus, the domain $D_{\hspace{-0.1cm}\tensor[_{\varphi}]{\hat{Q}}{^{3}_{\hspace{-0.1cm}phys}}(\tau)}$ of a single-particle s.a. extension $\tensor[_{\varphi}]{\hat{Q}}{^{3}_{\hspace{-0.1cm}phys}}(\tau)$ will be given by $D_{\tensor[_{\varphi}]{\hat{Q}}{^{3,+}_{\hspace{-0.4cm}phys}}(\tau)}\oplus D_{\tensor[_{\varphi}]{\hat{Q}}{^{3,-}_{\hspace{-0.4cm}phys}}(\tau)}$, its spectrum being fully defined through the spectra of the operators $\tensor[_{\varphi}]{\hat{Q}}{^{3,\pm}_{\hspace{-0.1cm}phys}}(\tau)$. 

To formally obtain the spectra of operators $\tensor[_{\varphi}]{\hat{Q}}{^{3}_{\hspace{-0.1cm}phys}}(\tau)$ it is necessary to verify if the solutions
\begin{equation}
\mathcal{V}^{z}_{(\xi);\tau}(\nu_{\pi}) = \frac{\mathcal{N}^{z}_{\xi}}{(\sec\nu_{\pi})^{3/2}}e^{im\tau\ln(\sec\nu_{\pi})}e^{-im\xi z \nu_{\pi}}
\begin{pmatrix}
\delta_{\xi +} \\
\delta_{\xi -}
\end{pmatrix} \nonumber
\end{equation}
of $\check{Q}^{3}_{phys}(\tau)\mathcal{V}^{z}_{(\xi);\tau}(\nu_{\pi}) = z\mathcal{V}^{z}_{(\xi);\tau}(\nu_{\pi})$ belong to $D_{\hspace{-0.1cm}\tensor[_{\varphi}]{\hat{Q}}{^{3}_{\hspace{-0.1cm}phys}}(\tau)}$. Since $\mathcal{V}^{z}_{(\xi);\tau}(\nu_{\pi}) \in D^{*}_{\hat{Q}^{3}_{phys}(\tau)}$, it remains to verify the consequences of the boundary condition \eqref{eq:Bound1}. Applying \eqref{eq:Bound1} to $\mathcal{V}^{z}_{(\xi);\tau}(\nu_{\pi})$ implies that $\mathcal{V}^{z}_{(\xi);\tau}(\nu_{\pi}) \in D_{\hspace{-0.1cm}\tensor[_{\varphi}]{\hat{Q}}{^{3}_{\hspace{-0.1cm}phys}}(\tau)}$ only for a discrete set of eigenvalues $z_{n}$ given by
\begin{equation}
z_{\varphi}^{n} = \frac{2}{m\uppi}\left\{\,\mathrm{arctan}\left[\frac{(1-\cos\varphi)}{\sin\varphi}\tanh\left(\frac{\uppi}{2}\right)\right] + n\pi\right\}, \nonumber
\end{equation}
with $n \in \mathbb{Z}$ and $-\pi/2 \leq \mathrm{arctan}(\alpha) \leq \pi/2$. Thus, the complete set of orthogonal eigenfunctions of $\tensor[_{\varphi}]{\hat{Q}}{^{3}_{\hspace{-0.1cm}phys}}(\tau)$ is given by the solutions
\begin{equation}
\mathcal{V}^{z^{n}_{\varphi}}_{(\xi);\tau}(\nu_{\pi}) = \frac{e^{im\tau\ln(\sec\nu_{\pi})}e^{-im\xi z^{n}_{\varphi} \nu_{\pi}}}{\uppi^{1/2}(\sec\nu_{\pi})^{3/2}}\begin{pmatrix}
\delta_{\xi +} \\
\delta_{\xi -}
\end{pmatrix}. \nonumber
\end{equation}

The discrete spectrum of the s.a. extensions $\tensor[_{\varphi}]{\hat{Q}}{^{3}_{\hspace{-0.1cm}phys}}(\tau)$ may seem unsatisfactory at first glance, since a continuous spectrum is expected for observables associated with the system's position. However, continuity can be recovered when the set of all s.a. extensions $\tensor[_{\varphi}]{\hat{Q}}{^{3}_{\hspace{-0.1cm}phys}}(\tau)$ is taken into account, since $z_{\uppi}^{n} = \lim_{\varphi \rightarrow -\uppi}z^{n+1}_{\varphi}$.

To finish the description of the s.a. extensions of $\hat{Q}^{3}_{phys}(\tau)$ it is necessary to verify the properties of the others operators in its CSCO. Then, the complete set of generalized orthogonal eigenfunctions of the s.a. CSCO of $\tensor[_{\varphi}]{\hat{Q}}{^{3}_{\hspace{-0.1cm}phys}}(\tau)$ is given by the states
\begin{widetext}
\begin{equation}
\psi^{z^n_{\varphi},\lambda,m_{z}}_{(\xi);\tau}(\bs{\pi}) = \frac{\sqrt{\sinh(\uppi\Lambda(\lambda))}}{2}\frac{|\Gamma(\frac{1}{2} + |m_{z}| + i\Lambda(\lambda))|}{(m\uppi)^{3/2}}\frac{e^{im\tau\ln(\sec\nu_{\pi})}e^{-im\xi z^{n}_{\varphi}\nu_{\pi}}e^{im_{z}\varphi_{\pi}}}{(\sec\nu_{\pi})^{3/2}}P^{-|m_{z}|}_{-\frac{1}{2} + i\Lambda(\lambda)}(\cosh\omega_{\pi})\begin{pmatrix}
\delta_{\xi +} \\
\delta_{\xi -}
\end{pmatrix}, \label{eq:EigenQ3}
\end{equation}
\end{widetext}
where $\Gamma(\cdot)$ is the Gamma function, $P^{-|m_{z}|}_{-\frac{1}{2} + i\Lambda(\lambda)}(\cosh\omega_{\pi})$ are associated conical functions, $\Lambda(\lambda) = \sqrt{-\frac{1}{4}-\lambda}$, $m_z \in \mathbb{Z}$ are the eigenvalues of the s.a. operator $\hat{J}^{12}_{phys}$ and $\lambda \in (-\infty,-\frac{1}{4}]$ are the values of the continuous spectrum of the s.a. extension of $\hat{O}_{phys}$.

The fact that $\hat{Q}^{3}_{phys}(\tau)$ has s.a. single-particle extensions suggests that the operators $\tensor[_{\varphi}]{\hat{Q}}{^{j}_{\hspace{-0.1cm}phys}}(\tau)$ should be interpreted as the observables associated with the system's proper-time parameterized position. However, some considerations indicate that this interpretation should be refuted, starting with the fact that the discrete spectrum of these operators does not allow a continuous description of the system's position by means of a single s.a. extension with fixed $\varphi$ parameter. A second problem is related to Hegerfeldt's paradox. Since the operator $\tensor[_{\varphi}]{\hat{Q}}{^{3}_{\hspace{-0.1cm}phys}}(\tau)$ is s.a., its eigenstates describe strictly localized states on the z-axis and, therefore, are subject to the causality violation predicted by Hegerfeldt.

\emph{Time and Position POVM -} As stated earlier, the s.a. extensions of $\hat{Q}_{phys}^{0}(\tau)$ do not define single-particle observables, a result that leads to two important consequences: (a) the perfect temporal localization of a state is only possible outside the single-particle framework and (b) states with a well defined energy sign will necessarily have a temporal incertitude when described by a classical observer.

In view of that, one may ask how a measurement associated to the operation $\check{Q}^{0}_{phys}(\tau)$ may be accommodated in a single-particle framework. To do so, it is necessary to observe that the projections $\hat{Q}^{0,\pm}_{phys}(\tau)$ of $\hat{Q}^{0}_{phys}(\tau)$ over the subspaces $\mathcal{H}_{phys}^{\pm}$ are essentially maximally symmetrical \cite{Gitman2012} and, therefore, their closures lead to maximally symmetrical operators $\underline{\hat{Q}^{0;\pm}_{phys}(\tau)}$ that can be associated with POVMs \cite{Egusquiza2008}. Thus, although a single-particle interpretation cannot be associated to the s.a. extensions $\tensor[_{\varphi}]{\hat{Q}}{^{0}_{\hspace{-0.1cm}phys}}(\tau)$, the same is not true for the maximally symmetrical operators $\underline{\hat{Q}^{0;\pm}_{phys}(\tau)}$. Therefore, the implementation of the time operator in the proper-time single-particle formalism must be given by a POVM rather than a s.a. operator.  In addition, since the domain boundary conditions of $\underline{\hat{Q}^{0}_{phys}(\tau)}$ do not mix components associated with distinct energy signs, one has that $\underline{\hat{Q}^{0}_{phys}(\tau)} = \underline{\hat{Q}^{0;+}_{phys}(\tau)} \oplus \underline{\hat{Q}^{0;-}_{phys}(\tau)}$ and the domain properties of the operators $\underline{\hat{Q}^{0;\pm}_{phys}(\tau)}$ coincide with those obtained earlier for $\underline{\hat{Q}^{0}_{phys}(\tau)}$.

The set of positive operators $\{\hat{E}_{\tau;\pm}(t)\}$ associated with the POVM defined by $\underline{\hat{Q}^{0;\pm}_{phys}(\tau)}$ can be obtained using Naimark's theorem \cite{Busch1995} and is given by
\begin{equation}
\hat{E}_{\tau;\pm}(t) = \sum_{l=0}^{\infty}\sum_{m_{z}=-l}^{l}\ket{\psi^{t,l,m_z}_{\tau;\pm}}\bra{\psi^{t,l,m_z}_{\tau;\pm}}, \nonumber
\end{equation}
with
\begin{equation}
\psi^{t,l,m_z}_{\tau;\pm}(\bs{\pi}) = \sqrt{\frac{m}{2\uppi}}\frac{Y^{l,m_{z}}(\Omega_{\pi})}{r_{\pi}^{3/2}}\hspace{-0.1cm}\left(\frac{r_{\pi}}{m}\right)^{im\tau}\hspace{-0.1cm}\left(\frac{r_{\pi}}{E_{r_{\pi}} + m}\right)^{\mp imt}.
\nonumber
\end{equation}
Thus, the probability of finding the system in a time interval $[t_1,t_2]$ for a state described by a density matrix $\rho$ is given by $P_{[t_1,t_2]} = \int_{t_{1}}^{t_{2}}dt \mathrm{Tr}(\rho\hat{E}_{\tau;\pm}(t))$, while the completeness relation in $\mathcal{H}_{phys}^{\pm}$ associated with the elements $\hat{E}_{\pm}(t)$ is written as $\int_{-\infty}^{\infty}dt \hat{E}_{\tau;\pm}(t) = \hat{I}^{\pm}_{phys}$. It is important to note that the elements $\hat{E}_{\tau;\pm}(t)$ for different values of $t$ are not orthogonal, since
\begin{equation}
\braket{\psi^{t,l,m_z}_{\tau;\pm}|\psi^{t',l',m'_z}_{\tau;\pm}} = \frac{\delta_{ll'}\delta_{m_{z}m_{z}'}}{2\uppi}\left[\uppi\delta(\Delta t) \pm \mathrm{P.V.}\left(\frac{i}{\Delta t}\right)\right], \nonumber
\end{equation}
where $\mathrm{P.V.}$ indicates the principal value and $\Delta t \equiv t - t'$. Thus, the strict temporal localization is not possible in the proper-time single-particle formalism and physically acceptable states will necessarily present a temporal uncertainty.

To verify how the imposition of the single-particle character over the $\check{Q}^{0}_{phys}(\tau)$ operation influences the system's spacial localization, the covariance of the quantities $\hat{Q}^{\mu}_{phys}(\tau)$, which follows from the commutation relations \eqref{eq:ASD}, needs to be taken into account. Denoting by $\hat{U}(\Lambda)$ the unitary representations \cite{Tung1985} of the Lorentz group generators $\hat{J}^{\mu\nu}_{phys}$ and assuming a pure z-axis boost $\Lambda$, the relations in \eqref{eq:ASD} imply that
\begin{equation}
\begin{aligned}
\hat{Q}^{3'}_{phys}(\tau) & \equiv \hat{U}^{\dagger}(\Lambda)\hat{Q}^{\mu}_{phys}(\tau)\hat{U}(\Lambda) \\ & = \tensor{\Lambda}{^{3}_{0}}\hat{Q}^{0}_{phys}(\tau) + \tensor{\Lambda}{^{3}_{3}}\hat{Q}^{3}_{phys}(\tau). 
\end{aligned} \label{eq:Lor1}
\end{equation}
Since the domain of $\hat{Q}^{3}_{phys}(\tau)$ cannot change from one reference frame to another, relationship \eqref{eq:Lor1} implies that one must have $D_{\hat{Q}^{3}_{phys}(\tau)} = D_{\hat{Q}^{0}_{phys}(\tau)}\cap D_{\hat{Q}^{3}_{phys}(\tau)}$ in order to respect the finite Lorentz covariance.  However, assuming that in subspace $\mathcal{H}_{phys}^{\xi}$ the time operator is given by $\underline{\hat{Q}^{0,\xi}_{phys}(\tau)}$, there are states in $D_{\tensor[_{\varphi}]{\hat{Q}}{^{3,\xi}_{\hspace{-0.4cm}phys}}(\tau)}$ that will not belong to  $D_{\underline{\hat{Q}^{0,\xi}_{phys}(\tau)}}$, since the boundary condition \eqref{eq:Bound1} imposed by $D_{\tensor[_{\varphi}]{\hat{Q}}{^{3,\xi}_{\hspace{-0.4cm}phys}}(\tau)}$ admits states that don't cancel out faster than $r_{\pi}^{-3/2}$ for $\nu_{\pi}\rightarrow\pm \uppi/2$, as is the case for the eigenstates \eqref{eq:EigenQ3}. Thus, the single-particle character together with finite Lorentz covariance rules out the possibility to use the s.a. extensions of $\hat{Q}_{phys}^{j}(\tau)$ as position operators.

To overcome the above problem, one can adopt  $\underline{\hat{Q}^{j,\pm}_{phys}(\tau)}$ as definitions of the single-particle position operator in each $\mathcal{H}_{phys}^{\pm}$ subspace, since the boundary conditions of these operators agree with those of $\underline{\hat{Q}^{0,\pm}_{phys}(\tau)}$. However, since $\underline{\hat{Q}^{j,\pm}_{phys}(\tau)}$ is symmetrical but not s.a., such a choice will imply in a concept of localization that cannot be associated with a projective measurement. In order to define a POVM associated with $\underline{\hat{Q}^{3,\xi}_{phys}(\tau)}$, one observe that the eigenstates $\mathcal{V}^{z^{n}_{\varphi}}_{\tau;\xi}(\tau)$ of $\tensor[_{\varphi}]{\hat{Q}}{^{3,\xi}_{\hspace{-0.4cm}phys}}(\tau)$ also serve as an improper basis for $D_{\underline{\hat{Q}^{3,\xi}_{phys}(\tau)}}$, since $D_{\underline{\hat{Q}^{3,\xi}_{phys}(\tau)}}  \subset D_{\tensor[_{\varphi}]{\hat{Q}}{^{3,\xi}_{\hspace{-0.4cm}phys}}(\tau)}$ and $\mathcal{V}^{z^{n}_{\varphi}}_{\tau;\xi}(\tau) \notin D_{\underline{\hat{Q}^{3,\xi}_{phys}(\tau)}}$.

Using the improper basis of states $\psi^{z^{n}_{\varphi},\lambda,m_{z}}_{\tau;\xi}(\bs{\pi})$, the operator $\underline{\hat{Q}^{3,\xi}_{phys}(\tau)}$ can be written as
\begin{equation}
\underline{\hat{Q}^{3,\xi}_{phys}(\tau)}\hspace{-0.05cm} = \hspace{-0.1cm}\sum_{n\in\mathbb{Z}}\sum_{m_{z}\in\mathbb{Z}}\int_{-\infty}^{-1/4}\hspace{-0.65cm}d\lambda \left(\hspace{-0.1cm}z^{0}_{\varphi} + \frac{2n}{m}\hspace{-0.1cm}\right)\hspace{-0.1cm}\ket{\psi^{z^{n}_{\varphi},\lambda,m_{z}}_{\tau;\xi}}\hspace{-0.1cm}\bra{\psi^{z^{n}_{\varphi},\lambda,m_{z}}_{\tau;\xi}}. \nonumber
\end{equation}
Since $D_{\underline{\hat{Q}^{3,\xi}_{phys}(\tau)}} \subset D_{\tensor[_{\varphi}]{\hat{Q}}{^{3,\xi}_{\hspace{-0.4cm}phys}}(\tau)}$ for all $\varphi\in(-\uppi,\uppi]$, the above decomposition can be done using the basis of eigenstate of any of the s.a. extensions $\tensor[_{\varphi}]{\hat{Q}}{^{3,\xi}_{\hspace{-0.4cm}phys}}(\tau)$ and, therefore, an integration in $z^{0}_{\varphi}\in[-1/m,1/m]$ allows to rewrite $\underline{\hat{Q}^{3,\xi}_{phys}(\tau)}$ as
\begin{equation}
\underline{\hat{Q}^{3,\xi}_{phys}(\tau)} = \int_{\mathbb{R}}\hspace{-0.1cm}\frac{m dz}{2}\hspace{-0.1cm}\sum_{m_{z}\in\mathbb{Z}}\int_{-\infty}^{-1/4}\hspace{-0.8cm}d\lambda \,z\ket{\psi^{z,\lambda,m_{z}}_{\tau;\xi}}\bra{\psi^{z,\lambda,m_{z}}_{\tau;\xi}}, \label{eq:DecSO}
\end{equation}
where $\psi^{z,\lambda,m_{z}}_{\tau;\xi}(\bs{\pi})$ is given by the $\xi$-component of \eqref{eq:EigenQ3} with $z_{\varphi}^{n}$ replaced by $z\in\mathbb{R}$. This continuous description of the spacial position is non-orthogonal, since
\begin{equation}
\braket{\psi^{z',\lambda',m_{z}'}_{\tau;\xi}|\psi^{z,\lambda,m_{z}}_{\tau;\xi}} = \delta_{m_{z}'m_{z}}\delta(\lambda'-\lambda)\mathrm{sinc}\hspace{-2pt}\left(\frac{m\uppi(z'-z)}{2}\right). \label{eq:NonOrt}
\end{equation}
However, the non-orthogonality decays with $[m(z'-z)]^{-1}$, i.e. it decreases with the inverse of the number of Compton wavelengths separating $z'$ from $z$, being relevant only for small values of $\Delta z$.

The above results, along with the fact that the identity $\hat{I}^{\xi}_{phys}$ in $\mathcal{H}^{\xi}_{phys}$ can be written as
\begin{equation}
\hat{I}^{\xi}_{phys} = \frac{m}{2}\int_{-\infty}^{\infty}dz\sum_{m_{z}\in\mathbb{Z}}\int_{-\infty}^{-1/4}d\lambda\ket{\psi^{z,\lambda,m_{z}}_{\tau;\xi}}\bra{\psi^{z,\lambda,m_{z}}_{\tau;\xi}}, \nonumber
\end{equation}
allows to introduce a POVM associated with the operator $\underline{\hat{Q}^{3,\xi}_{phys}(\tau)}$, the positive operators $\{\hat{E}_{\tau;\xi}(z)\}$ associated with that POVM being given by
\begin{equation}
\hat{E}_{\tau;\xi}(z) = \frac{m}{2}\sum_{m_{z}\in\mathbb{Z}}\int_{-\infty}^{-1/4}d\lambda\ket{\psi^{z,\lambda,m_{z}}_{\tau;\xi}}\bra{\psi^{z,\lambda,m_{z}}_{\tau;\xi}}. \label{eq:POVMpos}
\end{equation}
These operators satisfy $\int_{\mathbb{R}}dz\hat{E}_{\tau;\xi}(z) = \hat{I}^{\xi}_{phys}$ and the probability $P_{[z_1,z_2]}(\tau)$ of finding a state of density matrix $\rho$ in a spatial range $[z_1,z_2]$ for a proper-time $\tau$ is given by $P_{[z_1,z_2]}(\tau) = \int_{z_{1}}^{z_2}dz \mathrm{Tr}\left(\rho\hat{E}_{\tau;\pm}(z)\right)$.

\emph{Hegerfeldt's theorem and the POVM approach -} To conclude the description of the system's localization in terms of $\underline{\hat{Q}^{3,\xi}_{phys}(\tau)}$ it is necessary to verify that the POVM given in \eqref{eq:POVMpos} is not subject to the causality issues related to Hegerfeldt's paradox \cite{Hegerfeldt1974,Hegerfeldt1980,Hegerfeldt1985}. Since those results assert the causality violation for strictly localized states as well as exponentially bounded states, it must be verified that such states are not allowed by the domain $D_{\underline{\hat{Q}^{3,\pm}_{phys}(\tau)}}$.

For the proof of the nonexistence of strictly localized states, suppose that the state
\begin{equation}
\ket{\psi} = \int_{-\infty}^{-1/4}d\lambda\alpha(\lambda)\int_{\mathbb{R}}dz \Omega(z)\ket{\psi_{0;+}^{z,\lambda,0}} \nonumber
\end{equation}
is strictly localized at $\tau = 0$. Due to the non-orthogonality given in \eqref{eq:NonOrt}, the strictly localization condition consists in supposing that the probability amplitude
\begin{equation}
p_{0}(z') = \sqrt{\frac{m}{2}}\int_{\mathbb{R}}dz \Omega(z) \mathrm{sinc}\left(\frac{m\uppi}{2}(z'-z)\right), \nonumber
\end{equation}
has a compact support in an interval $\Delta z \subset \mathbb{R}$.

The compact support of $p_{0}(z')$ implies that its Fourier
\begin{equation}
\mathcal{F}_{p_0}(k) = \frac{1}{\sqrt{2\uppi}}\int_{\mathbb{R}}dz' p_0(z') e^{-i\uppi m k z'} \label{eq:transFFF}
\end{equation}
must be analytic in $\mathbb{R}$. Making the change of variables given by $u = m\uppi z$, the Fourier \eqref{eq:transFFF} may be rewritten as
\begin{equation}
\mathcal{F}_{p_0}(k) = \frac{\sqrt{2\uppi}}{(m\uppi)^{3/2}}\mathrm{rect}(k)\mathcal{F}_{\Omega}(k), \label{eq:transFF}
\end{equation}
where $\mathrm{rect}(k)$ is a rectangular function and $\mathcal{F}_{\Omega}(k) = (2\uppi)^{-1/2}\int_{\mathbb{R}}du\Omega(u)e^{-iku}$.

To verify the conditions imposed by $D_{\underline{Q^{3;+}_{phys}(0)}}$, the state $\ket{\psi}$ must be written in momentum basis. In this basis
\begin{equation}
\psi_{+}(\bs{\pi}) = \frac{A_{0}}{m\uppi^{3/2}}\frac{\mathcal{F}_{\Omega}(k)}{(\sec\nu_{\pi})^{3/2}}, \nonumber
\end{equation}
where $A_{0}$ is a $\omega_{\pi}$-dependent factor and $k = \nu_{\pi}/\uppi$. From the conditions in the domain $D_{\underline{Q^{3;+}_{phys}(0)}}$ it results that $\mathcal{F}_{\Omega}(k)$ must be zero for $k = \pm 1/2$ and must belong to $L^{2}((-1/2,1/2),dk)$, besides being differentiable.

The properties obtained for $\mathcal{F}_{\Omega}(k)$ imply that the Fourier $\mathcal{F}_{p_0}(k)$ must be differentiable and have compact support in $[-1/2,1/2]\subset\mathbb{R}$, since $\mathcal{F}_{p_0}(k) \propto \mathrm{rect}(k)\mathcal{F}_{\Omega}(k)$. Therefore, the function $\mathcal{F}_{p_0}(k)$ cannot be analytic in $\mathbb{R}$ and $p_{0}(z')$ cannot have compact support, which demonstrates the nonexistence of strictly localized states with respect to the localization definition associated to $\underline{\hat{Q}^{3,\xi}_{phys}(\tau)}$.

The proof of the nonexistence of states that are compatible with $\underline{\hat{Q}^{3,\xi}_{phys}(\tau)}$ and have a probability amplitude $p_0(z')$ with tails bounded by an exponential decay $e^{-A|z|}$ follows the same reasoning presented above. In this case, the exponential behavior of the tails of $p_0(z')$ would imply the analyticity of the Fourier $\mathcal{F}_{p_0}(k)$ for $|\mathrm{Im}(k)| < A$. However, the domain $D_{\underline{Q^{3; +}_{phys}(0)}}$ implies that over $\mathbb{R}$ the function $\mathcal{F}_{p_0}(k)$ must have compact support in $[-1/2,1/2]$ and, therefore, the condition of analyticity in $|\mathrm{Im}(k)| < A$ cannot be satisfied, leading to the conclusion of the nonexistence of states with bounded exponential decay.

\emph{Additional remarks -} The proper-time parameterization has a fundamental character since it does not depend on the properties of an external observer and, therefore, corresponds to an intrinsic approach to the problem of localization in RQM. Physically, this approach amounts to state that the system's time would be observed as classical only if it were possible to define an observer from a comoving quantum frame as those proposed in \cite{Giacomini2019n}.

The impossibility of defining strictly localized states in a single-particle approach is in agreement with the idea that such localization would involve energies that would lead to a regime in which the phenomena of creation and annihilation of particles could no longer be disregarded. As obtained in \eqref{eq:NonOrt}, the distance between two orthogonal position in the z-axis is at least of two Compton wavelength, a range that is in agreement with what is expected from a regime with fixed number of particles as RQM.

It is worth emphasizing that the reported results lead to a new path to address the issue of localization in the context of RQM, further potential questions to be investigated including the proper definition of the relativistic spin operator, the connection with the quantum comoving frames proposed in \cite{Giacomini2019n} and the relation between the proposed POVMs and the usual measurements parameterized by quantities of the classical observer.

\emph{Acknowledgments -} The authors are thankful for the support provided by Brazilian agencies CAPES (PROCAD2013), CNPq (\#459339/2014-1, \#312723/2018-0), FAPEG (PRONEX \#201710267000503, PRONEM \#201710267000540) and the Instituto Nacional de Ciência e Tecnologia - Informação Quântica (INCT-IQ).

\bibliography{bib/Bibliografia,bib/bibtiqr,bib/Livros}

\end{document}